\documentclass[final,1p,times]{elsarticle}
\usepackage{graphics}
\usepackage{epsfig}

\usepackage{amssymb}
\begin{document}

\begin{frontmatter}



\title{Lattice Simulations near the Semimetal-Insulator Phase Transition of Graphene}

\author[label1]{Wesley Armour}
\author[label2]{Simon Hands}
\author[label3]{Costas Strouthos\corref{cor1}}
\address[label1]{Diamond Light Source, Harwell Campus, Didcot, Oxfordshire OX11 0DE, United Kingdom}
\address[label2]{Department of Physics, Swansea University, Singleton Park SA2 8PP, Swansea, United Kingdom}
\address[label3]{Department of Mechanical Engineering, 91 Aglanzias Avenue, University of Cyprus, Nicosia 1678, Cyprus}
\cortext[cor1]{Corresponding author. email:strouthos@ucy.ac.cy, tel.:+357 22894510, fax: +357 22892254}

\begin{abstract}
We present results from Monte Carlo simulations of a three dimensional fermion field
theory which can be derived from a model of graphene in which electrons interact
via a screened Coulomb potential. 
For our simulations we employ lattice gauge theory methods used in elementary particle 
physics. We show that the theory undergoes a second order phase transition and we provide
estimates for the critical exponents.  
The estimated value of the physical critical coupling implies that graphene in vacuum is
an insulator. We also present the first results for the quasiparticle dispersion relation.

\end{abstract}


\end{frontmatter}


\section{Introduction}
\label{intro}
There has been considerable recent interest in graphene (single layer graphite) sparked by its
discovery and subsequent experimental study \cite{Novoselov:2005kj}. 
The remarkable properties of graphene have been suspected for many years \cite{Semenoff:1984dq}.
In brief, for a carbon monolayer having one mobile
electron per atom,
a simple tight-binding model shows that the spectrum of low-energy excitations
exhibits a linear dispersion relation centred on zeros located at the six
corners of the first Brillouin
zone (see e.g. \cite{Gusynin2007}). Using a linear transformation among
the fields at two independent zeros it is possible to recast the Hamiltonian
in Dirac form with $N_f=2$ flavors of
four-component spinor $\psi$ - the counting of the massless
degrees of freedom coming from two
carbon atoms per unit cell times two zeros per zone times
two physical spin components per electron. Electron propagation in the graphene
layer is thus relativistic, albeit at a Fermi velocity $v_F\approx 10^6m/s$. 
The implications for the high mobility of the resulting charge carriers 
is the source of the current excitement. The electronic properties of graphene
are discussed in detail in a recent review paper \cite{CastroNeto}.

Although the semimetallic properties of graphene on a substrate are well-known,
it remains unclear whether graphene in vacuum undergoes a phase transition from
the semimetal phase to an insulating phase.
A suffuciently strong Coulomb interaction 
may lead to a condensation of electron and hole pairs thus turning the conducting phase into a 
gapped insulator at a critical value of fermion flavors $N_{fc}$. Recent estimates of $N_{fc}$ 
have been obtained by: self-consistent solutions of 
Schwinger-Dyson equations
yielding $N_{fc}=2.55$ \cite{Khveshchenko}; renormalization group treatment of 
radiatively induced four-fermion interactions yielding $N_{fc}=2.03$ \cite{Drut2007}; 
and lattice simulations 
yielding $N_{fc}=4.8(2)$ \cite{Hands2008}. 
Furtermore, a lattice simulation  
of graphene based on a four dimensional gauge invariant action with fermions propagating in
two spatial dimensions predicted that freely suspended graphene is an insulator
\cite{Drut2009,Drut2009a}. The results
of lattice simulations of a Thirring-like model pertinent to graphene presented 
in this paper are consistent with this observation.

\section{The Model}
\label{model}
While the above considerations apply quite generally, a realistic model of
graphene must
incorporate interactions between charge carriers. One such model, due to Son
\cite{Son:2007ja}, has $N_f$ massless fermion flavors propagating in the
plane, but interacting via an
instantaneous $3d$ Coulomb interaction. In Euclidean metric and static gauge
$\partial_0A_0=0$ the action reads
\begin{eqnarray}
\label{eq:SonModel}
S_1  &=& \sum_{a=1}^{N_f}\int dx_0d^2x(\bar\psi_a\gamma_0\partial_0\psi_a
+v_F\bar\psi_a\vec\gamma.\vec\nabla\psi_a
\nonumber \\
&+& iV\bar\psi_a\gamma_0\psi_a) 
+ {1\over{2e^2}}\int dx_0d^3x(\partial_i V)^2,
\label{eq:model}
\end{eqnarray}
where $e$ is the electron charge, $V\equiv A_0$ is the electrostatic potential,
 and the $4\times4$ Dirac matrices satisfy
$\{\gamma_\mu,\gamma_\nu\}=2\delta_{\mu\nu}$, $\mu=0,\dots,3$.
In our notation
$\vec x$ is a vector in the $2d$ plane while the index $i$ runs over all three
spatial directions.
In the large-$N_f$ limit the resummed
$V$ propagator becomes \cite{Son:2007ja}
\begin{equation}
\label{eq:D1}
D_1(p)= (D_0^{-1}(p)-\Pi(p))^{-1}
=\left({{2\vert\vec p\vert}\over e^2}+{N_f\over8}{{\vert\vec
p\vert^2}\over{(p^2)^{1\over2}}}\right)^{-1},
\label{eq:D_gr}
\end{equation}
where $p^2\equiv p_0^2+v_F^2\vert\vec p\vert^2$. $D_0$ 
corresponds to the classical propagation of $V$ and $\Pi$ 
comes from a vacuum polarisation fermion-antifermion loop.
In either the strong
coupling or large-$N_f$ limits $D_1(p)$ is thus dominated by
quantum corrections - the relative importance of the original Coulomb interaction being
governed by a parameter $\lambda\equiv\vert\Pi/D_0\vert_{p_0=0}$. 
Chiral symmetry breaking due to the spontaneous condensation of particle - hole
pairs is signalled by an order parameter $\langle\bar\psi\psi\rangle\not=0$.
Physically the most important outcome is the generation of a gap in the fermion
spectrum, implying the model is an insulator.

The physics of eq.~(\ref{eq:SonModel}) is very reminiscent of the three dimensional Thirring model, 
which
is analytically tractable at large-$N_f$, but exhibits spontaneous chiral symmetry breaking
at small $N_f$ and strong coupling \cite{DelDebbio:1997,DelDebbio:1999xg}.
Arguably the Thirring model is the simplest field theory of fermions requiring a
computational solution.
The location of the phase transition at $N_f=N_{fc}$ in the strong coupling
limit has recently been determined by lattice simulations to be
$N_{fc}=6.6(1)$~\cite{Christofi:2007ye}. The apparent similarity of the two
systems has led us to propose a Thirring-like model pertinent to graphene \cite{Hands2008}, 
with Lagrangian
\begin{equation}
\label{eq:model2}
S_2= \sum_{a=1}^{N_f}\int dx_0d^2x  \left[ \bar\psi_a\gamma_\mu\partial_\mu\psi_a
+iV\bar\psi_a\gamma_0\psi_a+{1\over{2g^2}}V^2 \right].
\end{equation}
As for eq.~(\ref{eq:model}) we assume a large-$N_f$ limit to
estimate the dominant vacuum polarisation correction; the resultant propagator
for $V$ is
\begin{equation}
\label{eq:D2}
D_2(p)
=\left({1\over g^2}+{N_f\over8}{{\vert\vec
p\vert^2}\over{(p^2)^{1\over2}}}\right)^{-1}.
\label{eq:D_Th}
\end{equation}
In the strong-coupling or large-$N_f$ limits,
$D_2$ coincides with $D_1$,
implying that the fermion interactions are equivalent.
We thus expect predictions made with the model eq.~(\ref{eq:model2}), and in
particular critical behavior such as the
value of $N_{fc}=4.8(2)$ \cite{Hands2008},
to be generally valid for Son's model eq.~(\ref{eq:model}) in the limit of large
$\lambda$. 

\section{Lattice Simulation Results}
\label{result}

\begin{figure}[t]
%\begin{minipage}{17pc}
\begin{center}
\includegraphics[width=19pc]{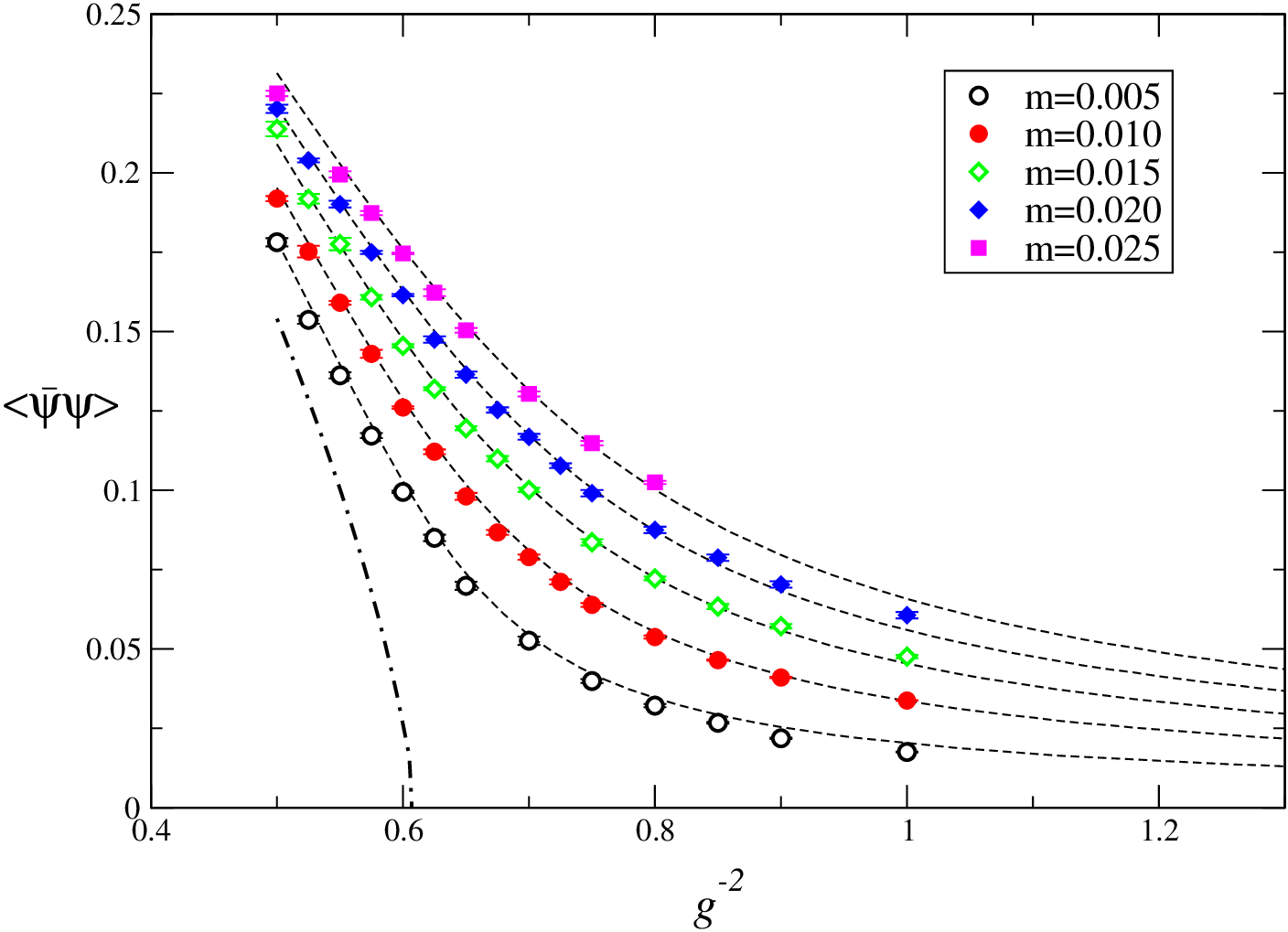}
\caption{\label{fig:eos} $\langle\bar\psi\psi\rangle$ versus $g^{-2}$ with fits
to  eq.~(\ref{eq:truncEoS}).}
\end{center}
%\end{minipage}\hspace{2pc}
\end{figure}
%\vspace{0.2in}

In this section we present results from simulations  
with $N_f=2$ which corresponds to real graphene. 
Details for the lattice action and the algorithm employed for our simulations can be found in
\cite{Hands2008}.
The anisotropic nature of the model dynamics results in 
the systematic effects due to finite temporal extend $L_t$ being 
more important than those due to finite 
spatial extend $L_s$. By comparing the values of the chiral condensate from simulations on different 
lattice sizes we found that the finite size effects on $24^{2}\times48$ are within 
statistical errors for the parameter ranges $m_0=0.005$, $0.525\! \leq\! g^{-2}\! \leq\! 0.65$ and
 $m_0=0.010,...,0.025$, 
$0.525 \leq g^{-2} \leq 0.70$ (where $m_0$ is the fermion bare mass). 
The chiral condensate as a function of $m_0$ and 
$(g_c^{-2}-g^{-2})$ near a continuous phase transition scales 
according to an equation of state given by
\begin{equation}
\label{eq:truncEoS}
m_0=A(g^{-2}-g_c^{-2})\langle\bar\psi\psi\rangle^{\delta-\frac{1}{\beta}}
+B\langle\bar\psi\psi\rangle^\delta
\end{equation}
We used eq.~(\ref{eq:truncEoS}) to curve-fit our data and extracted $g_c^{-2}=0.607(5)$, 
$\delta=2.72(7)$, and $\beta=0.68(2)$. The data and the fitting curves including the $m_0=0$ curve
are shown in Fig.~\ref{fig:eos}. Our estimates of the critical exponents are
clearly distinct from their strong coupling values $\delta=5.5(3)$ and 
$\beta=0.22(2)$ \cite{Hands2008}.
We also note that the exponent $\delta$ is larger than the
estimate $\delta=2.26(6)$ of Drut and L\"ahde \cite{Drut2009a}.

\begin{figure}[h]
\begin{center}
\includegraphics[width=19pc]{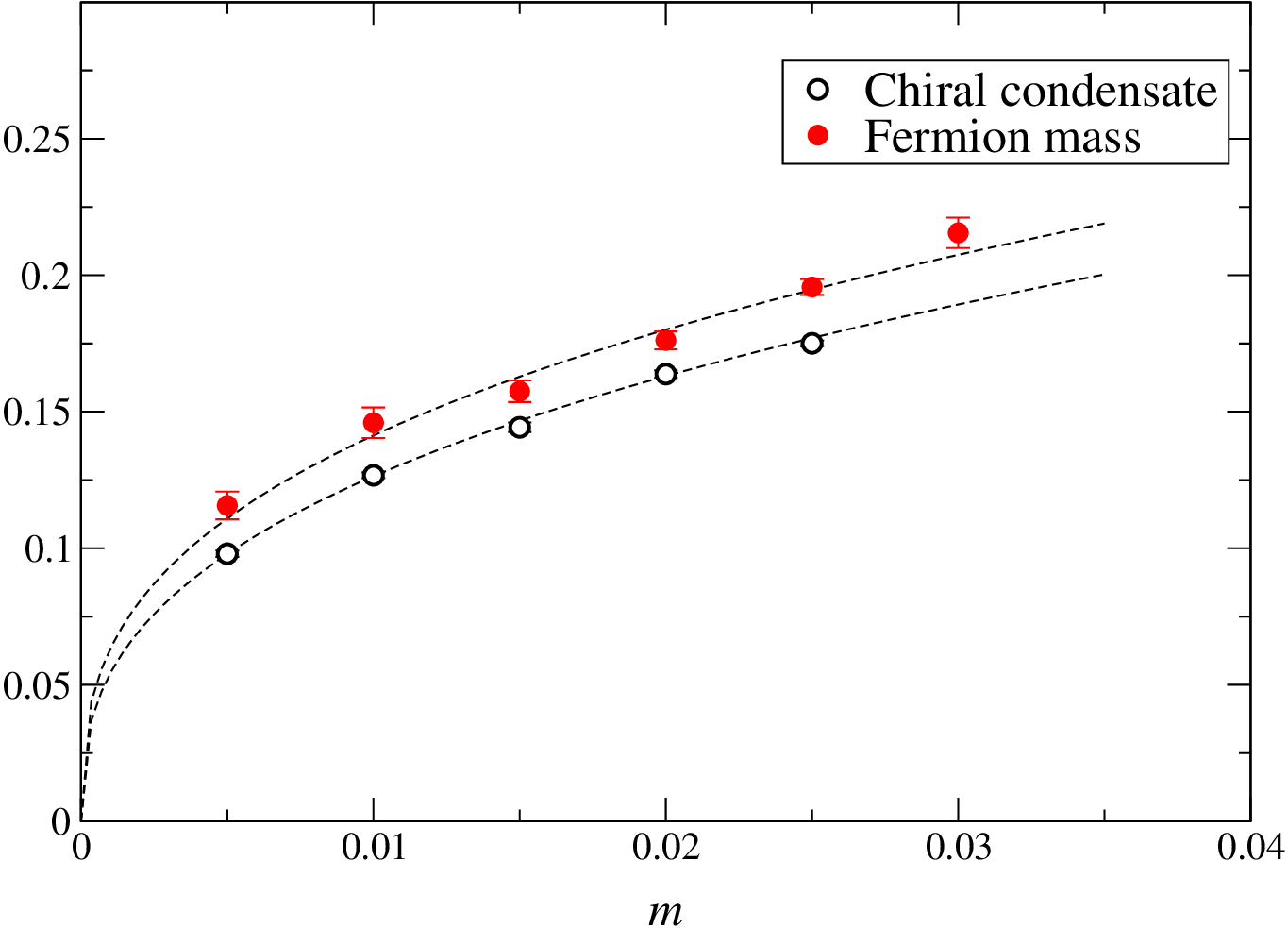}
\caption{\label{fig:crit}$\langle\bar\psi\psi\rangle$ and $m_f$ fits to eq.~(\ref{eq:nonzerom}) 
at $\beta=0.60$.}
\end{center}
%\end{minipage}
\end{figure}

The fermion dynamical mass $m_f$ is an inverse time-like correlation length of the system.
At the critical coupling, $\langle\bar\psi\psi\rangle$ and $m_f$ obey the
following scaling relations for $m_0 \neq 0$: 
\begin{equation}
\label{eq:nonzerom}
\langle\bar\psi\psi\rangle = a_1m_0^{1/\delta}, \hspace{0.3in} m_f=a_2m_0^{\nu_t/\beta\delta}. 
\end{equation}
By fitting our data at $\beta=0.60$ (which is close to
the critical coupling $g^{-2}=0.607(5)$)  
to eq.~(\ref{eq:nonzerom}) we obtain $\delta=2.81(2)$ and $\nu_t=0.80(6)$. 
This value of $\delta$ is consistent with $\delta=2.72(7)$ obtained from the
equation of state. In Fig.~\ref{fig:crit} we present the data and the fitting functions.

For systems with
anisotropic interactions the correlation lengths in the time and space directions
may scale with different exponents $\nu_t$ and $\nu_s$ respectively. The dynamic critical
exponent $z \equiv \nu_t/\nu_s$ describes relative scaling of time and space length
scales and governs the quasiparticle dispersion relation $E \sim p^z$
for $p \rightarrow 0$.
In $d$ spacetime dimensions the exponents $\nu_t$ and $\nu_s$ are related to $\beta$
and $\delta$ via a generalized hyperscaling relation \cite{Hornreich}:
\begin{equation}
\label{eq:hyper}
\nu_t + (d-1)\nu_s = \beta(\delta+1)
\end{equation}
From eq.~(\ref{eq:hyper}) we obtain $\nu_s=0.90(5)$ and $z=0.89(8)$. This value of $z$ 
is close to the strong coupling and large-$N_f$ $z \approx 0.8$ estimate of Son \cite{Son:2007ja}.
A value of $z<1$ implies that the quasiparticle excitations are stable,
because energy momentum conservation forbids their decay into two or more quasiparticles.

Before we decide whether our model predicts that graphene in vacuum is a conductor  
or an insulator we need an estimate of the physical critical coupling $\lambda_c$. 
First we calculate 
the renormalized critical coupling $g_{cR}^{-2}$. 
We must note however, that the vacuum
polarisation calculation leading to eqs.~(\ref{eq:D1}) and (\ref{eq:D2})
does not go through in quite the same way for the lattice regularised model;
rather, there is an additive correction which is 
momentum independent and ultraviolet-divergent: 
\begin{equation}
\Pi^{latt}(p)=\Pi^{cont}(p)+g^2J(m),
\end{equation}
where $J(m)$ comes from incomplete cancellation of a lattice tadpole diagram
~\cite{DelDebbio:1997}. This extra divergence not
present in the continuum treatments can be absorbed by a wavefunction
renormalisation of $V$ and a coupling constant renormalisation
\begin{equation}
\label{eq:gr}
g_R^2={g^2\over{1-g^2J(m)}}.
\label{eq:gR}
\end{equation}
In the large-$N_f$ limit
we thus expect to find the strong coupling limit of the lattice model at
$g_R^2\to\infty$ implying $g^2\to g^2_{\rm lim}=J^{-1}(m)$.
For $g^2>g^2_{\rm lim}$ $D^{latt}(p)$ becomes negative, and $S_{latt}$ no longer describes a
unitary theory. 
For small-$N_f$ we identify $g^2_{\rm lim}$ with the value of the coupling where 
the chiral condensate has its maximum value \cite{Hands2008}. For $N_f=2$,  
$g^2_{\rm lim}=0.30$ \cite{Hands2008} and from eq.~(\ref{eq:gr}) we obtain 
$g_{cR}=3.23(4)$. Next, we match $D_2^{\rm latt}(p)$ to $D_1(p)$
at a reference momentum $p_0=0$, $|\vec{p}|=\frac{\pi}{2}$, 
so that the propagators are equal at a distance of roughly one lattice spacing (for
shorter distances $D_2^{latt} > D_1$ and vice-versa) yielding the condition
\begin{equation}
\lambda_c=\frac{\pi g^2_{cR}}{4}=2.54(2).
\end{equation}
This value of $\lambda_c$ is larger than the value $\lambda_c \approx1.25$ 
calculated for graphene on an $\rm SiO_2$ substrate \cite{Son:2007ja}, but smaller 
than the value $\lambda_0 \approx 3.4$ based on a substrate with unit dielectric constant. 
Our result thus suggests freely-suspended graphene is an insulator. 
Drut and L\"ahde obtained from simulations of a gauge invariant action of graphene 
$\lambda_c=1.67(2)$, which also implies that graphene in vacuum is an insulator \cite{Drut2009}.

\begin{figure}[t] 
\label{fig:dispersion}
\begin{center} 
\includegraphics[width=19pc]{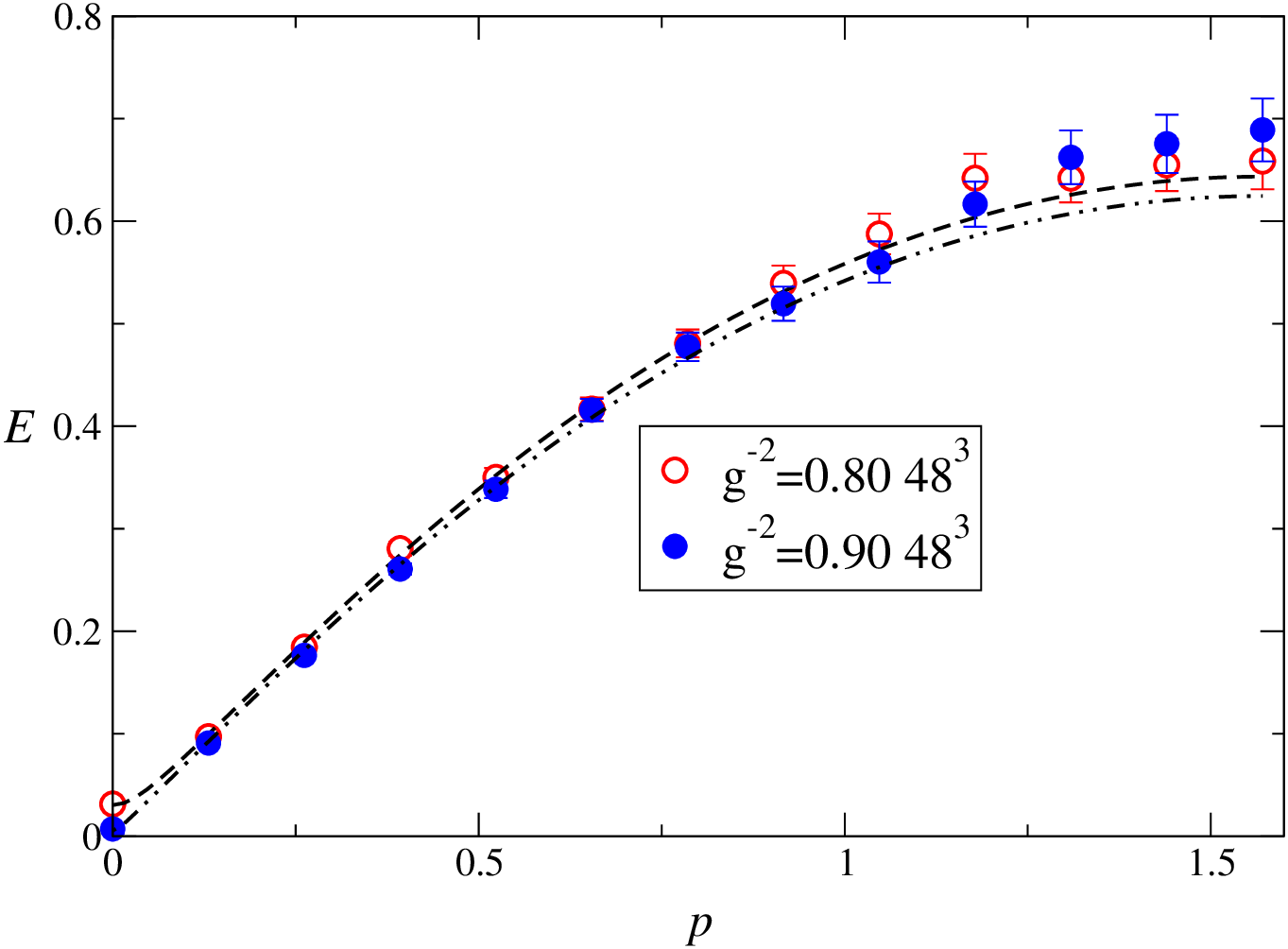}
\caption{ Dispersion relation at $g^{-2}=0.8, 0.9$ and $m_0=0.005$ on a $48^3$ lattice 
with fits to eq.~(\ref{dispersion}).}
\end{center}
\end{figure}

\begin{figure}[ht]
\begin{center}
\includegraphics[width=20pc]{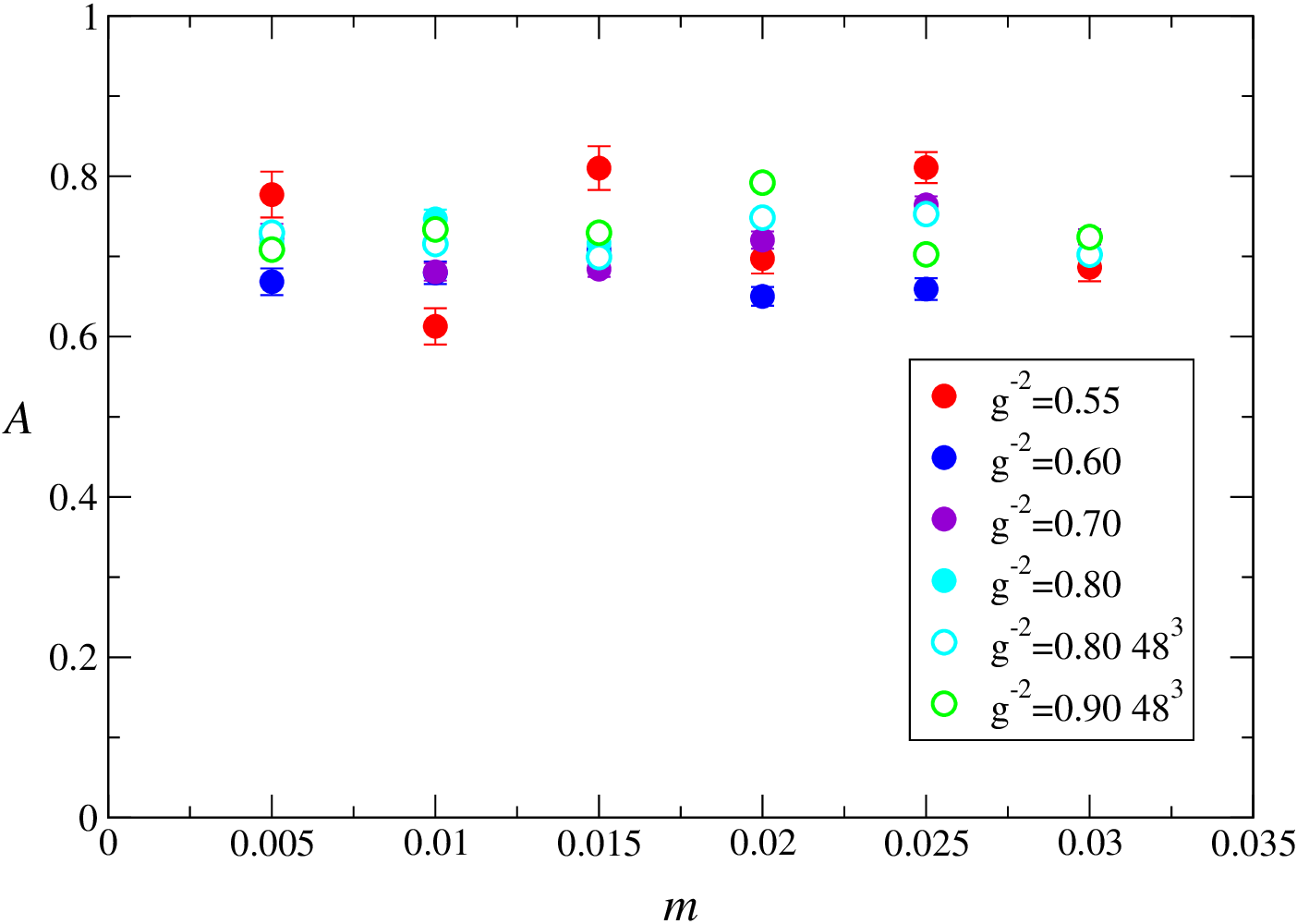}
\caption{\label{fig:A} Values of $A$ versus $m_0$ extracted from fits with eq.~(\ref{dispersion}) 
on $32^2\times 48$ and $48^3$ lattices.}
\end{center}
\end{figure}

A major advantage of our formulation is that it permits a relatively straight-forward analysis 
of the quasiparticle dispersion relation.
Here we present our first results from 
simulations on $32^2\times48$ and $48^3$ lattices. We fitted the fermion energy as 
a function of momentum $p$
in the $x$ direction to the following 
discretized free field dispersion relation which assumes $z=1$:
\begin{equation}
\label{dispersion}
E(p)=A\sinh^{-1}\bigl(\sqrt{\sin^2p+M^2}\bigr),
\end{equation}
where $AM$ is the quasiparticle dynamical mass $m_f$ and
$A \sim \lim_{p \rightarrow 0} \frac{dE}{dp}$ is the Fermi velocity. Equation (\ref{dispersion}) 
fitted the data relatively well, providing no evidence for $z \neq 1$. An example of fits
to data generated with $g^{-2}=0.8, 0.9$ and $m_0=0.005$ on a $48^3$ lattice is shown in 
Fig.~\ref{fig:dispersion}.
In Fig.~\ref{fig:A} we show that within the accuracy 
of our analysis the parameter $A$ 
is independent of both $m_0$ and $g^{-2}$ with a value roughly $A \approx 0.7$ 
suggesting that the free field value 
of $v_F$ is renormalized. 

\section{Summary and Conclusions}
\label{summary}
In this paper we studied a $(2+1)d$ Thirring-like model pertinent to graphene 
based on eq.~(\ref{eq:model2}).
We performed Monte Carlo simulations using lattice gauge theory techniques and showed that the model 
undergoes a second order phase transition with critical exponents $\delta=2.72(7)$, $\beta=0.68(2)$, 
and $\nu_t=0.80(6)$. Using a generalized hyperscaling relation valid 
for theories with anisotropic interactions
we obtained $\nu_s=0.90(5)$ and $z=0.89(8)$. Our preliminary $z<1$ estimate implies 
that the quasiparticles
are stable based on energy-momentum conservation. 
The value of the physical critical coupling $\lambda_c=2.54(2)$ implies that 
freely-suspended graphene is an insulator in agreement with \cite{Drut2009}. We also presented the 
first results for the quasiparticle dispersion relation and within the accuracy of our simulations
we showed that the free field Fermi velocity is renormalized. Simulations with higher resolution in 
momentum on larger spatial lattices are currently in progress for a more accurate study of 
the dispersion relation.

\end{document}